\documentstyle[pra,aps]{revtex}

\begin{document}
\draft\onecolumn
\baselineskip=12pt

\title{\bf  Electron-radiation interaction in 
a Penning trap:\\
beyond the dipole approximation}

\author{Ana M. Martins$^{\dag}$, Stefano Mancini$^{\ddag}$, 
and Paolo Tombesi$^{\ddag}$}

\address{${\dag}$ Centro de Fisica de Plasmas, 
Instituto Superior T\'ecnico,\\
P-1096 Lisboa Codex, Portugal\\
$\ddag$ Dipartimento di Matematica e Fisica and 
Istituto Nazionale per la Fisica della Materia,\\
Universit\`a di Camerino, 
I-62032 Camerino, Italy}

\date{Received: \today}

\maketitle\widetext

\begin{abstract}
We investigate the physics of a single trapped electron 
interacting with a radiation field without
the dipole approximation. This gives new physical insights 
in the so-called geonium theory.
\end{abstract}

\pacs{PACS number(s): 03.65.Bz, 42.50.Dv, 12.20.-m}

\widetext

\section{Introduction}

Simple systems like single electron or ion provide a useful 
tool to investigate the
fundamental laws of nature.
Hence in the past decades there has been increasing interest 
on trapping phenomena
\cite{paul}.
It is now routinely
possible to trap a single ion \cite{cohen}, 
and it would allow us to study 
QED (Quantum Electro-Dynamics) also
when the trapped ion interacts with a radiation mode.
On the other hand the electron stored in a Penning trap 
\cite{pen} permitted accurate
measurements too \cite{van}.
This system has been called "geonium atom" since 
it resembles an hydrogen atom where
indeed the binding for the electron is to an external 
apparatus residing on the earth \cite{BG}.
The geonium system was recently studied to
implement some interesting quantum optics situations, like QND 
(Quantum Non-Demolition) measurements
\cite{irene} and the generation of nonclassical states \cite{rapid}.
 
Here, just after one hundred years from the 
electron discovery, 
we would show new interesting quantum features arising 
on a trapped 
electron interacting 
with the radiation field, when no dipole approximation is made.

It is well known that in the geonium system \cite{BG} the 
motion of the electron can be separated into 
three independent harmonic motions: the axial, 
cyclotron and magnetron. 
On the other hand, 
it is also well established that entangled systems are
extremely interesting for many purposes.

In the present work we propose a way of coupling the three 
harmonic oscillators of the geonium
system by simply superposing a radiation 
field to the trapping fields. 
More concretely, we show that when the
trapped electron oscillates in a standing wave field, 
there could be linear, or nonlinear coupling among
the axial motion and the other ones; 
although, in particular we will only consider 
the axial-cyclotron interaction. 
Hence, we shall present the more immediate consequenses 
of such an entanglement, like indirect measurements on the 
cyclotron 
mode, then we shall investigate the 
generation of nonclassical features.
Moreover, the analysis in all cases will be performed 
by taking into account the environmental effects as well.

The paper is organized as follow:
Section II is devoted to the description of the model.
The first order coupling (linear) between axial and cyclotron 
motion is considered in Section III, while in Section IV
the second order coupling is discussed.
In Section V we further discuss 
the possibility of generating nonclassical states.
Finally, we present conclusions.

\section{The Model}

The geonium system consists \cite{BG} of an electron of 
charge $e$ and 
mass $m_0$ moving in 
a uniform 
magnetic field ${\bf B}$, along the positive $z$ axis, 
and a static 
quadrupole potential
\begin{equation}\label{V}
{\hat V}=V_0\frac{{\hat x}^2+{\hat y}^2-2{\hat z}^2}{4d^2}\,,
\end{equation}
where $d$ characterizes the dimension of the trap and $V_0$ 
is the potential applied to the trap
electrodes \cite{BG}.

In this work, in addition to the usual trapping fields, 
we embed the 
trapped electron
in a radiation field of vector potential ${\hat{\bf A}}_{ext}$.
To simplify our presentation, we assume the {\it a priori} 
knowledge 
of the electron spin \cite{kells}. Neglecting all 
spin-related 
terms, the Hamiltonian for
the trapped electron can then be written as the 
quantum counterpart
of the classical Hamiltonian
\begin{equation}\label{Hinit}
{\hat H}=\frac{1}{2m_0}\left[{\hat{\bf p}}
-\frac{e}{c}{\hat{\bf A}}\right]^2
+e{\hat V}\,,
\end{equation}
where $c$ is the speed of light, and
\begin{equation}\label{A}
{\hat{\bf A}}=\frac{1}{2}{\hat{\bf r}}\wedge{\bf B}
+{\hat{\bf A}}_{ext}\,,
\end{equation}
where ${\hat{\bf r}}\equiv({\hat x},{\hat y},{\hat z})$ 
and ${\hat{\bf p}}\equiv({\hat p}_x,{\hat p}_y,{\hat p}_z)$ 
are respectively the position and the
conjugate momentum operators of the electron.

The motion of the electron in absence of the external field 
${\hat{\bf A}}_{ext}$ 
is the result of the motion of three
harmonic oscillators \cite{BG}: the cyclotron, the axial and 
the magnetron, which are well separated in the
energy scale (GHz, MHz and kHz respectively). This can be 
easily understood by
introducing the ladder operators
\begin{eqnarray}\label{lad}
{\hat a}_z&=&\sqrt{\frac{m_0\omega_z}{2\hbar}}\,{\hat z}
+i\,\sqrt{\frac{1}{2\hbar m_0\omega_z}}\,{\hat p_z}\\
{\hat a}_c&=&\frac{1}{2}\left[\sqrt{\frac{m_0\omega_c}{2\hbar}}
({\hat x}-i{\hat y})
+\sqrt{\frac{2}{\hbar m_0\omega_c}}({\hat p_y}
+i{\hat p_x})\right]\\
{\hat a}_m&=&\frac{1}{2}\left[\sqrt{\frac{m_0\omega_c}{2\hbar}}
({\hat x}+i{\hat y})
-\sqrt{\frac{2}{\hbar m_0\omega_c}}({\hat p_y}
-i{\hat p_x})\right]
\end{eqnarray}
where the indexes $z$, $c$ and $m$ stand for axial, cyclotron
and magnetron respectively.
The above operators obey the commutation relation 
$[{\hat a}_{i},{\hat a}^{\dag}_{i}]=1$, $i=z,\,c,\,m$.
The angular frequencies are given by
\begin{equation}\label{freq}
\omega_z=\sqrt{\frac{eV_0}{m_0cd^2}}\,;\quad
\omega_c=\frac{eB}{m_0c}\,;\quad
\omega_m\approx\frac{\omega_z^2}{2\omega_c}\,.
\end{equation}
So, when ${\hat{\bf A}}_{ext}=0$, the Hamiltonian 
(\ref{Hinit}) simply reduces to
\begin{equation}\label{Hfree}
{\hat H}=\hbar\omega_z\left({\hat a}_z^{\dag}{\hat a}_z
+\frac{1}{2}\right)
+\hbar\omega_c\left({\hat a}_c^{\dag}{\hat a}_c
+\frac{1}{2}\right)
-\hbar\omega_m\left({\hat a}_m^{\dag}{\hat a}_m
+\frac{1}{2}\right)\,.
\end{equation}

Instead, when the external radiation field is a 
standing wave along 
the $z$ direction (with frequency $\Omega$ and wave vector $k$)
and circularly polarized in the $x-y$ plane \cite{gatan}, we have
\begin{equation}\label{Aext}
{\hat{\bf A}}_{ext}=\left(
-i\left[\alpha e^{i\Omega t}-\alpha^*e^{-i\Omega t}\right]
\cos(k{\hat z}+\phi),
\left[\alpha e^{i\Omega t}+\alpha^*e^{-i\Omega t}\right]
\cos(k{\hat z}+\phi),
0\right)\,.
\end{equation}

In such a case, and for frequencies $\Omega$ close to 
$\omega_c$, we can neglect the slow
magnetron motion, and the Hamiltonian (\ref{Hinit})
becomes
\begin{eqnarray}\label{Hnodip}
{\hat H}&=&\hbar\omega_z\left({\hat a}_z^{\dag}{\hat a}_z
+\frac{1}{2}\right)
+\hbar\omega_c\left({\hat a}_c^{\dag}{\hat a}_c
+\frac{1}{2}\right)\nonumber\\
&+&\hbar\left[\epsilon^*{\hat a}_ce^{i\Omega t}
+\epsilon{\hat a}_c^{\dag}e^{-i\Omega t}\right]
\cos(k{\hat z}+\phi)
+\hbar\chi\cos^2(k{\hat z}+\phi)
\end{eqnarray}
where
\begin{equation}\label{eps}
\epsilon=|\epsilon|e^{i\varphi}
=\left(\frac{2e^3B}{\hbar m_0^2c^3}\right)^{1/2}
\alpha;\quad \chi=\frac{e^2}{\hbar m_0c^2}|\alpha|^2\,,
\end{equation}
and the phase $\varphi$ is the phase of the applied 
radiation field (i.e. ${\rm arg}\,\alpha$).
The other phase $\phi$ defines the position of 
the center of the axial motion with respect
to the standing wave. 
The third and fourth terms in the right hand side of the 
Hamiltonian
(\ref{Hnodip}) describe the nonlinear interaction between 
the trapped electron and the standing wave
which gives rise to a coupling between the axial 
and the cyclotron
motion, whose effect will be analyzed in the next sections.
In the usual Penning traps the quantity $k\langle{\hat z}\rangle$
can reach values up to
$\approx 0.1$ \cite{BG}, when 
$\Omega\approx\omega_c$.
This leads us to explore the physics beyond the usual 
dipole approximation 
for the cosine term in Eq. (\ref{Hnodip}).
The cosine factor $\cos(k{\hat z}+\phi)$ can be split as 
\begin{equation}\label{cos}
\cos(k{\hat z}+\phi)=\cos\phi\cos(k{\hat z})
-\sin\phi\sin(k{\hat z})\,,
\end{equation}
and two typical situations corresponding to 
$\phi=0$ and $\phi=\pi/2$ can be easily exploited.
By making the usual dipole approximation these two cases  
correspond to 
a mere driving term on the cyclotron motion ($\phi=0$) 
or to no effect at all ($\phi=\pi/2$).

In the following Sections the behaviour 
of the trapped electron in these two
paradigmatic limits is studied. All the other possible values of 
$\phi$  will give rise to combinations
of these two cases and can be easily studied.

We further note that the last term in (\ref{Hnodip}) 
can be neglected since the parameters
(\ref{eps}) are such that 
$\chi/|\epsilon|\approx|\epsilon|/\omega_c$.

\section{The Case of $\phi=\pi/2$}

In this section we consider the case $\phi=\pi/2$.
Developing $\sin(k{\hat z})$ in power series and keeping 
only the first order term 
we can approximate the Hamiltonian (\ref{Hnodip}) by
\begin{eqnarray}\label{Hsin}
{\hat H}&=&\hbar\omega_z\left({\hat a}_z^{\dag}{\hat a}_z
+\frac{1}{2}\right)
+\hbar\omega_c\left({\hat a}_c^{\dag}{\hat a}_c
+\frac{1}{2}\right)\nonumber\\
&+&\hbar\left[\epsilon^*{\hat a}_ce^{i\Omega t}
+\epsilon{\hat a}_c^{\dag}e^{-i\Omega t}\right]k{\hat z}\,.
\end{eqnarray}
In the case of perfect resonance, $\Omega=\omega_c$, 
and in a frame rotating at that angular
frequency we get the solution
\begin{eqnarray}
{\hat z}(t)&=&\left[{\hat z}(0)
-|\epsilon|k{\hat X}_{\varphi}\right]
\cos(\omega_zt)
+\frac{1}{m\omega_z}{\hat p}_z(0)\sin(\omega_zt)
+|\epsilon|k{\hat X}_{\varphi}\label{zt} \\
{\hat p}_z(t)&=&{\hat p}_z(0)\cos(\omega_zt)
-m\omega_z\left[{\hat z}(0)
-\sqrt{2}|\epsilon|k{\hat X}_{\varphi}\right]
\sin(\omega_zt)\label{pzt} 
\end{eqnarray}
where we have introduced the cyclotron quadrature
\begin{equation}\label{cquad}
{\hat X}_{\varphi}=\frac{{\hat a}_ce^{i\varphi}
+{\hat a}_c^{\dag}e^{-i\varphi}}{\sqrt{2}}\,.
\end{equation}

Equation (\ref{pzt}) suggests us an indirect way to 
determine the 
probability distribution 
for the cyclotronic quadrature, ${\cal P}(X_{\varphi})$.
We recall that in the geonium system the measurements 
are performed only 
on the axial degree of freedom due to the non existence 
of good detectors 
in the microwave regime.
The oscillating charged particle induces alternating 
image charges on the 
electrodes, which in turn cause an oscillating current 
to flow through an 
external circuit.
The current will be
proportional to the axial momentum ${\hat p}_z$, hence a 
measurement of this current will also give
the value of the quadrature ${\hat X}_{\varphi}$.
Measurements when the standing wave is `off' should be done 
preventively to set the initial conditions.
Then, repeated measurements lead to the desired statistics
${\cal P}(X_{\varphi})$. 

If the procedure is further repeated for several values 
of the phase $\varphi$, we obtain
the set of marginal probabilities ${\cal P}(X,\varphi)$,
which allows the tomographic imaging of
the quantum state of the cyclotron mode \cite{cyctom}.

We now consider the effects of the thermal damping 
through the resistance of the external 
circuit connected with the measurement apparatus. 
In such a case the equations of motion for the axial 
degree of freedom become
\begin{eqnarray}\label{eqzth}
\frac{d {\hat z}}{dt}&=&\frac{{\hat p}_z}{m_0}\,,\\
\frac{d {\hat p}_z}{dt}&=&-\omega_z^2 m_0 {\hat z}
-\frac{\gamma_z}{m_0}{\hat p}_z
-\sqrt{2}\hbar k |\epsilon|{\hat X}_{\varphi}+{\xi}\,,
\end{eqnarray}
where the noise term $\xi(t)$ is that of Johnson noise 
with expectation values
$\langle\xi(t)\rangle=0$, and $\langle\xi(t)\xi(t')\rangle
=2\gamma_zK_BT\delta(t-t')$, the
damping constant $\gamma_z$ is proportional to the 
readout resistor, $K_B$ is the Boltzmann
constant and $T$ the equilibrium temperature.

By using the Fourier transforms, we immediately obtain
\begin{equation}\label{pztilde}
{\tilde p}_z(\omega)=\frac{\sqrt{2}\hbar k 
|\epsilon|{\tilde X}_{\varphi}(\omega)-{\tilde\xi}(\omega)}
{\omega^2-\omega_z^2-i\omega\gamma_z/m_0}\,,
\end{equation}
hence the correlation
\begin{equation}\label{pzcorr}
\langle{\tilde p}_z(\omega) {\tilde p}_z(-\omega) \rangle=
\frac{2(\hbar k|\epsilon|)^2\langle {\tilde X}_{\varphi}
(\omega){\tilde X}_{\varphi}
(-\omega)\rangle
+\langle{\tilde\xi}(\omega){\tilde\xi}(-\omega)\rangle}
{\left|\omega^2-\omega_z^2-i\omega\gamma_z/m_0\right|^2}\,.
\end{equation}
Eq. (\ref{pzcorr}) imposes some limits to the observability 
of nonclassical effects on the cyclotron
motion; in fact the added thermal noise should be much less 
than the cyclotron vacuum noise for the
chosen frequency, i.e.
$\gamma_zK_BT<<(\hbar k|\epsilon|)^2$.

\section{The case of $\phi=0$}

Let us now consider the case of $\phi=0$, and keeping only 
terms up to the second order in $k{\hat
z}$, the Hamiltonian (\ref{Hnodip}) reduces to
\begin{eqnarray}\label{Hcos1}
{\hat H}&=&\hbar\omega_z\left({\hat a}_z^{\dag}{\hat a}_z
+\frac{1}{2}\right)
+\hbar\omega_c\left({\hat a}_c^{\dag}{\hat a}_c
+\frac{1}{2}\right)\nonumber\\
&+&\hbar\left[\epsilon^*{\hat a}_ce^{i\Omega t}
+\epsilon{\hat a}_c^{\dag}e^{-i\Omega t}\right]\left[1
-\frac{k^2{\hat z}^2}{2}\right]\,,
\end{eqnarray}
which clearly shows the nonlinear coupling as a consequence of 
the higher order expansion with respect to the case of the  
previous Section.

Let us study the general case including losses.
The latter are present in the axial degree of freedom once 
the connection with the external circuit is established, 
as pointed out in 
the previous Section. Instead,
the noise on the cyclotron degree of freedom could 
arise e.g. from radiative damping (though it can be 
strongly reduced
with an appropriate trap geometry).

Hence, by starting from the Hamiltonian (\ref{Hcos1}), 
we obtain the 
following Quantum Stochastic Differential Equations
\begin{eqnarray}
\frac{d{\hat a}_c}{dt}&=&-i\Delta{\hat a}_c
-\frac{\gamma_c}{2}{\hat a}_c
-i\epsilon(1-\kappa^2{\hat Z}^2)+\sqrt{\gamma_c}\,
{\hat a}_c^{in}\,,\label{lange1}\\
\frac{d{\hat a}^{\dag}_c}{dt}&=&i\Delta {\hat a}^{\dag}_c
-\frac{\gamma_c}{2}{\hat a}^{\dag}_c
+i\epsilon^*(1-\kappa^2{\hat Z}^2)+\sqrt{\gamma_c}\,
[{\hat a}_c^{in}]^{\dag}\,,\label{lange2}\\
\frac{d{\hat Z}}{dt}&=&\omega_z{\hat P}_z\,,
\label{lange3}\\
\frac{d{\hat P}_z}{dt}&=&-\omega_z{\hat Z}
+2\kappa^2(\epsilon^*{\hat a}_c
+\epsilon {\hat a}^{\dag}_c){\hat Z}
+f-\frac{\gamma_z}{m_0}{\hat P}_z
-\,{\Xi}\,,\label{lange4}
\end{eqnarray}
where $\Delta=\omega_c-\Omega$, $f$ is a driving 
term for the axial motion, $\gamma_c$ the cyclotron damping  
constant, and 
${\hat a}_c^{in}$, ${\Xi}$ are the noise 
terms (we shall consider the situation where only
the vacuum contributes to the cyclotron noise).
We have introduced the scaled variables
${\hat Z}=\sqrt{m_0\omega_z/\hbar}\,{\hat z}$, 
${\hat P}_z=\sqrt{1/\hbar m_0\omega_z}\,{\hat p}_z$, 
${\Xi}=\sqrt{1/\hbar m_0\omega_z}\,{\xi}$,
and $\kappa^2=\hbar k^2/2m_0\omega_z$.
From the Eq. (\ref{lange4}) we can see that the cyclotron 
quadrature causes a shift of the resonant frequency 
of the axial motion,
so its indirect measurement results feasible.

The system of equations (\ref{lange1}-\ref{lange4}) 
can be linearized around the steady state \cite{miwa}.
The stationary values ${\overline\alpha}_c$, 
${\overline Z}$ and
${\overline P_Z}$ can be obtained from the 
following equations
\begin{eqnarray}\label{steady}
0&=&-\left(\frac{\gamma_c}{2}
+i\Delta\right){\overline\alpha}_c-i\epsilon
(1-\kappa^2{\overline Z}^2)\,,\\
0&=&-\left(\frac{\gamma_c}{2}
-i\Delta\right){\overline\alpha}^*_c+i\epsilon^*
(1-\kappa^2{\overline Z}^2)\,,\\
0&=&\omega_z{\overline P}_Z\,,\\
0&=&-\left[\omega_z-2\kappa^2(
\epsilon^*{\overline\alpha}_c
+\epsilon{\overline\alpha}^*_c)\right]
{\overline Z}+f\,.
\end{eqnarray}
The linearized system is then
\begin{eqnarray}\label{linsys}
\frac{d}{dt}\left(
\begin{array}{cc}
{\hat a}_c\\
{\hat a}^{\dag}_c\\
{\hat Z}\\
{\hat P}_z
\end{array}\right)
={\bf M}\left(
\begin{array}{cc}
{\hat a}_c\\
{\hat a}^{\dag}_c\\
{\hat Z}\\
{\hat P}_z
\end{array}\right)
+\left(
\begin{array}{cc}
\sqrt{\gamma_c}\,{\hat a}_c^{in}\\
\sqrt{\gamma_c}\,\left[{\hat a}_c^{in}\right]^{\dag}\\
0\\
-{\Xi}
\end{array}\right)\,,
\end{eqnarray}
where now the operators indicate the quantum fluctuations
with respect to the steady state, and
\begin{eqnarray}\label{Adrift}
{\bf M}=\left(
\begin{array}{cccc}
-\left(\frac{\gamma_c}{2}
+i\Delta\right)&0&2i\epsilon\kappa^2{\overline Z}&0\\
0&-\left(\frac{\gamma_c}{2}
-i\Delta\right)&-2i\epsilon^*\kappa^2{\overline Z}&0\\
0&0&0&\omega_z\\
-2\epsilon^*\kappa^2{\overline Z}&
-2\epsilon\kappa^2{\overline Z}
&-\omega_z+2\kappa^2(\epsilon^*{\overline\alpha}_c
+\epsilon{\overline\alpha}^*_c)
&-\frac{\gamma_z}{m_0}
\end{array}\right)\,.
\end{eqnarray}

The spectral matrix can be calculated as
\begin{equation}\label{spec}
{\bf S}(\omega)=(i\omega{\bf I}-{\bf M})^{-1}{\bf D}
(-i\omega{\bf I}-{\bf M}^T)^{-1}\,,
\end{equation}
where ${\bf I}$ is the four by four identity matrix, 
${\bf M}^T$ means the 
transposed, and
\begin{eqnarray}\label{Ddiff}
{\bf D}=\left(\begin{array}{cccc}
0&\gamma_c&0&0\\
0&0&0&0\\
0&0&0&0\\
0&0&0&\frac{\gamma_z}{m_0}N_{th}
\end{array}\right)\,,
\end{eqnarray}
with $N_{th}=K_BT/\hbar\omega_z$ the 
number of thermal excitations.

The momentum correlation for the axial motion 
will be ${\bf S}_{44}$; this quantity is
plotted in Fig.1. The dashed line represents the 
resonance in absence of 
coupling and the solid line the resonance in presence 
of it. The 
separation between peaks is proportional to the 
cyclotron quadrature 
amplitude. So, it gives us an indirect value of 
that cyclotron observable.  

Furthermore, the variance for 
the amplitude cyclotron quadrature is given by integrating 
the quantity
${\bf S}_{11}+{\bf S}_{22}+{\bf S}_{12}+{\bf S}_{21}$, 
and it is plotted in Fig.2 (dashed line).
The same figure also shows the variance for the 
orthogonal quadrature 
(solid line).
It can be seen that the system exhibits 
squeezing effects depending on the detuning.
It is worth noting that such effects are not much sensitive
to thermal noise.

The stability of the system, for the values of parameters used,  
is checked through the signs of the eigenvalues of the matrix
${\bf M}$.

In this Section and in the previous one we have shown 
that the terms 
beyond the dipole approximation could play an important 
role and should 
not be neglected abruptely. As a matter of fact we have 
presented a 
variety of effects (see e.g. Figs.1 and 2) that 
could be measured in common Penning traps.

To go further, in the following, we shall explore 
other possibilities.

\section{Nonclassical States}

We now demonstrate the
generation of nonclassical effects due 
to the nonlinearity induced by Hamiltonian (\ref{Hcos1}).

\subsection{The central resonance}

If we tune the standing wave at frequency $\Omega=\omega_c$, 
and pass to the
interaction picture, the Hamiltonian (\ref{Hcos1}) 
simply becomes
\begin{equation}\label{Hcos}
{\hat H}=\sqrt{2}\hbar|\epsilon|{\hat X}_c\left[1
-\kappa^2\left({\hat a}_z^{\dag}{\hat a}_z
+\frac{1}{2}\right)\right]\,,
\end{equation}
where we have disregarded the rapidly oscillating 
terms ${\hat a}_z^{\dag}e^{-2i\omega_zt}$
and ${\hat a}_ze^{2i\omega_zt}$ (i.e. we made the 
Rotating Wave Approximation). 

Starting from initial coherent states for both modes
\begin{equation}\label{inista}
|\Psi(0)\rangle=|\alpha\rangle_c\otimes|\beta\rangle_z\,,
\end{equation} 
we obtain from the Hamiltonian (\ref{Hcos}) 
the following state at the time $t$
\begin{equation}\label{Psit}
|\Psi(t)\rangle=e^{-|\beta|^2/2}\sum^{\infty}_{n=0}
\frac{\beta^n}{\sqrt{n!}}
|\theta_nt\rangle_c\otimes|n\rangle_z\,,
\end{equation}
where
$\theta_n=
i\epsilon\kappa^2n$. In writing the state (\ref{Psit})  we have 
disregarded, for the sake of simplicity, the quantity 
$\alpha-i\epsilon(1-\kappa^2/2)t$, which is common to each 
cyclotron component
(this corresponds to an overall displacement in 
the cyclotron phase space).

Therefore, the electron motion evolves classically 
as a mixture of 
coherent states. 
Thus, during the evolution, no nonclassical states of the 
electron are generated. However, because
of the entanglement between the cyclotron and the axial 
degrees of freedom, 
it is possible to generate nonclassical states
of the cyclotron motion by performing 
conditional measurements 
on the axial degree of freedom.
In particular, a measurement of the axial current 
corresponds to the projection onto an eigenstate
$|p_z\rangle$ of the axial momentum
\begin{equation}\label{cyccat}
|\Psi(t)\rangle_{after}={\cal N}\sum^{\infty}_{n=0}
\left[
e^{-|\beta|^2/2}\frac{\beta^n}{\sqrt{n!}}
\langle p_z|n\rangle_z\right]
|\theta_nt\rangle_c\otimes|p_z\rangle\,,
\end{equation}
where ${\cal N}$ is a normalization constant and 
$\langle p_z|n\rangle_z$ 
are the harmonic oscillator wave functions in the 
momentum space.
It is immediately seen from the above expression, 
that after 
the measurement the system is left in a
superposition of coherent cyclotronic states which could 
have nonclassical features. 
Whenever the number of coherent states that are being 
superposed 
is small, the states are known as Schr\"odinger 
cat states \cite{sch}.

It is worth noting that the separation between the 
superposed coherent states is given by
$|\epsilon|\kappa^2t$, and therefore it can be made 
truly macroscopic 
emphasizing the nonclassicality (by simply requiring that 
$|\epsilon|\kappa^2t > 1$).
However, one has to be careful when satisfying the above
condition, since it also implies a  strong excitation 
of the cyclotron motion (the overall displacement 
that has been 
disregarded), which in turn could give rise to
instabilities or even, the loss of 
the particle over the trap's walls.

The Wigner function of the cyclotron state generated 
by conditional measurement can be written as 
\begin{eqnarray}\label{Wig}
W(Q,P)={\cal N}^{\,2}\,\sum_{m,n}c^*_mc_n\exp\Bigg[
&-&Q^2-P^2
-\frac{|\zeta_m|^2}{2}-\frac{|\zeta_n|^2}{2}\nonumber\\
&+&\sqrt{2}Q(\zeta_n+\zeta^*_m)-\sqrt{2}iP(\zeta_n
-\zeta^*_m)-\zeta_n\zeta_m^*\Bigg]\,,
\end{eqnarray}
where the variable $Q$, $P$ are associated to 
the quadratures ${\hat X}_{\varphi=0}$ and
${\hat X}_{\varphi=\pi/2}$ respectively, and
\begin{eqnarray}\label{parameters}
c_n&=&e^{-|\beta|^2/2}\frac{\beta^n}{\sqrt{n!}}
\langle p_z|n\rangle_z\,,\\
\zeta_n&=&\theta_nt\,.
\end{eqnarray}

In Fig. 3 we present the Wigner function of the 
cyclotron state generated by conditional measurement
on the axial degree of freedom. The negative parts 
and several oscillations show the
nonclassicality of such a state.

We have considered the measurement process conditioning 
the cyclotron state as instantaneous,
however, it always takes a finite time during which the 
system undergoes the back-action of the
measurement apparatus. To take into account these 
effects we should adopt a precise Hamiltonian
model describing the measurement of the observable 
${\hat p}_z$.
Nevertheless, from a phenomenological point of view, 
we can model the measurement process, performed on the state 
$\hat\rho(t)=|\Psi(t)\rangle\langle\Psi(t)|$,
during a time $\tau$, as the transition
\begin{equation}\label{red}
{\hat\rho}(t)\rightarrow {\rm Tr}_z\left[{\hat\rho}
(t+\tau)\,|p_z\rangle\langle p_z|\right]\,,
\end{equation}
where ${\hat\rho}(t+\tau)$ is obtained from 
${\hat\rho}(t)$ through free evolution 
in a thermal bath (representing the back-action 
of the measurement apparatus on the system), while
the projector indicates the output resulting 
at the end of the measurement \cite{proj}.

To evaluate the effects of the measurement on the
cyclotron state we should calculate the reduced 
density operator (r.h.s. of Eq.(\ref{red})).
Its corresponding Wigner function
is derived in Appendix A as
\begin{eqnarray}\label{Wth3}
W(Q,P,\tau)&=&{\cal N}^{\,2}\,\sum_{m,n}
\frac{\beta^m}{m!}
\frac{(\beta^*)^n}{n!}2^{-(m+n)/2}\,I_{m,n}
\exp\Bigg[
-Q^2-P^2\nonumber\\
&-&\frac{|\zeta_m|^2}{2}-\frac{|\zeta_n|^2}{2}
+\sqrt{2}Q(\zeta_m+\zeta^*_n)-\sqrt{2}iP(\zeta_m
-\zeta^*_n)-\zeta_m\zeta_n^*\Bigg]\,,
\end{eqnarray}
where
\begin{eqnarray}\label{Imn}
I_{m,n}&=&2^nm!\int dv\,\exp\left\{-\left[
e^{-2\Gamma\tau}+2N_{th}(1-e^{-2\Gamma\tau})\right]
v^2-2iP_zv\right\}\nonumber\\
&\times&\left(-ve^{-\Gamma\tau}\right)^{n-m}
L^{n-m}_m\left(2v^2e^{-2\Gamma\tau}\right)\,;
\quad\quad n>m\,,
\end{eqnarray}
with $L_m^n$ the associated Laguerre polynomials,
and $\Gamma=\gamma_z/m_0$ the effective axial 
damping constant.

The Wigner function (\ref{Wth3}) is  plotted in 
Fig.4 and shows the deleterious effects of finite 
time measurement on the nonclassical state represented in
Fig.3. Of course, these effects strongly depend  on the 
number of thermal excitations
$N_{th}$ as well. 

Once the cat states are generated 
by the conditional measurements, it would 
be possible to detect them by using indirect measurements 
as proposed in the previous Sections. 

\subsection{The sideband resonance}

We now return to the Hamiltonian (\ref{Hcos1}) to 
consider another resonance,  in this case 
$\Omega=(\omega_c-2\omega_z)-\delta$, where $\delta$ 
is a small detuning 
(i.e. $\delta<<\omega_z$) introduced for convenience. 
In a frame rotating at frequency $\omega_c-\delta$, 
we then have
\begin{equation}\label{Htri}
{\hat H}=\hbar\delta {\hat a}^{\dag}_c{\hat a}_c
-\hbar\frac{|\epsilon|\kappa^2}{2}
\left({\hat a}_c{\hat a}_z^{\dag 2}e^{-i\varphi}
+{\hat a}_c^{\dag}{\hat a}_z^2e^{i\varphi}
\right)\,.
\end{equation}
This is a trilinear Hamiltonian analogous to that 
studied in 
nonlinear optical processes like parametric oscillation 
or second harmonic generation \cite{miwa}. 

The equations of motion are 
\begin{eqnarray}\label{eqtri}
\frac{d {\hat a}_c}{dt}&=&-i\delta {\hat a}_c
+i\frac{1}{2}|\epsilon|\kappa^2{\hat a}_z^2\,,\\
\frac{d{\hat a}_z}{dt}&=&i|\epsilon|
\kappa^2{\hat a}^{\dag}_z{\hat a}_c\,;
\end{eqnarray}
and, by adiabatic elimination of the cyclotron mode, 
we get
\begin{equation}\label{adieq}
\frac{d{\hat a}_z}{dt}
=i\frac{|\epsilon|^2\kappa^4}{2\delta}
{\hat a}^{\dag}_z{\hat a}_z^2\,.
\end{equation}
This equation corresponds to an effective 
Hamiltonian for the axial motion of the type
\begin{equation}\label{Heff}
{\hat H}_{eff}=-\hbar
\frac{|\epsilon|^2\kappa^4}{4\delta}
({\hat a}^{\dag}_z)^2{\hat a}_z^2\,,
\end{equation}
which shows a well known Kerr-type nonlinearity. 
Hence, we should expect nonclassical effects, such as 
Schr\"odinger cat states,
when one starts from the initial conditions (\ref{inista}),
also in the axial mode.
In fact, the evolved axial state can be written as
\begin{equation}\label{psiz}
|\psi(t)\rangle_z=\exp\left[iG\left(
({\hat a}^{\dag}_z{\hat a}_z)^2
-{\hat a}^{\dag}_z{\hat a}_z\right)t\right]
|\beta\rangle_z\,,\quad
\quad G=
\frac{|\epsilon|^2\kappa^4}{4\delta}\,.
\end{equation}
It is easy to show that after a time $t=\pi/(2G)$ 
the initial coherent state evolves into a cat state
of the type discussed in Ref. \cite{YS}
\begin{equation}\label{cat}
|\psi(t=\pi/(2G))\rangle=\frac{1}{\sqrt{2}}\left[
e^{-i\pi/4}|-i\beta\rangle+e^{i\pi/4}|
i\beta\rangle\right]\,.
\end{equation}
That state shows interference in the phase space 
which could be detected by 
measuring an appropriate quadrature. Therefore, 
by adjusting the initial conditions we
may exploit the axial momentum measurement to 
see such interference.
The Wigner function of the state (\ref{cat}) results
\begin{eqnarray}\label{W10}
W(Z,P_z)&=&\frac{1}{\pi}e^{-|\beta|^2-Z^2-P_z^2}
\nonumber\\
&\times&\Bigg\{e^{-|\beta|^2}\cosh\left[2\sqrt{2}
P_z\,{\rm Re}(\beta)-2\sqrt{2}Z\,{\rm
Im}(\beta)\right]\nonumber\\ &+&e^{|\beta|^2}\sin
\left[2\sqrt{2}P_z\,{\rm Im}(\beta)
+2\sqrt{2}Z\,{\rm
Re}(\beta)\right]\Bigg\}\,,
\end{eqnarray}
and is represented in Fig.5. 
The fact that only two coherent states are being superposed 
is evident from 
the two hills besides the central structure, 
differentely  from the 
situation of Fig.3 where more components 
contributes to the cat state.

Of course, we should deal again with the problem 
of measurement, whose  
process renders the system open, 
hence, the dissipation
tends to wash out the nonclassical effects.
To evaluate this phenomenon we assume to switch off 
the nonlinearity at the time of cat generation,
and a subsequent free evolution of the axial 
degree of freedom in a thermal bath,
representing the effects of the external readout 
circuit.
If the latter takes a time $\tau$, we have
(see Appendix B)
\begin{equation}\label{Wsideth}
W(Z,P_z,\tau)=\frac{1}{2}e^{-|\beta|^2
+\beta^2/2+\beta^{*2}/2}
\left\{e^{2{\rm Im}(\beta)^2}\left[I_1+I_2\right]
-ie^{-2{\rm Re}(\beta)^2}\left[I_3-I_4\right]\right\}\,,
\end{equation}
where 
\begin{equation}\label{4int}
I_i=\frac{2}{\pi\sqrt{4{\cal A}{\cal B}-{\cal C}^2}}
\exp\left[\frac{{\cal B}{\cal D}_i^2
+{\cal C}{\cal D}_i{\cal E}_i+{\cal
A}{\cal E}_i^2}{4{\cal A}{\cal B}-{\cal C}^2}\right]\,;
\quad i=1,2,3,4,
\end{equation}
with
\begin{eqnarray}\label{ABC}
{\cal A}&=&\frac{1}{\Gamma^2}(e^{-\Gamma\tau}-1)^2+1\\
&+&2\frac{N_{th}}{\Gamma^2}(1-e^{-2\Gamma\tau})
-8\frac{N_{th}}{\Gamma^2}(1-e^{-\Gamma\tau})
+4\frac{N_{th}}{\Gamma}\tau\,;\nonumber\\
{\cal B}&=&e^{-2\Gamma\tau}+2N_{th}(1-e^{-2\Gamma\tau})\,;\\
{\cal C}&=&-\frac{2}{\Gamma}e^{-\Gamma\tau}
(e^{-\Gamma\tau}-1)
-4\frac{N_{th}}{\Gamma}(1-e^{-2\Gamma\tau})
+8\frac{N_{th}}{\Gamma}(1-e^{-\Gamma\tau})\,;
\end{eqnarray}
and
\begin{eqnarray}\label{DE}
{\cal D}_{1\atop 2}&=&\mp 2\sqrt{2}i{\rm Im}
(\beta)\mp\frac{\sqrt{2}}{\Gamma}i{\rm
Re}(\beta)e^{-\Gamma\tau}
\pm\frac{\sqrt{2}}{\Gamma}i{\rm Re}(\beta)+2iZ\,;\\
{\cal D}_{3\atop 4}&=&\mp 2\sqrt{2}{\rm Re}(\beta)
\pm\frac{\sqrt{2}}{\Gamma}{\rm
Im}(\beta)e^{-\Gamma\tau}
\mp\frac{\sqrt{2}}{\Gamma}{\rm Im}(\beta)+2iZ\,;\\
{\cal E}_{1\atop 2}&=&\mp 2\sqrt{2}i{\rm Re}(\beta)
e^{-\Gamma\tau}-2iP_z\,;\\
{\cal E}_{3\atop 4}&=&\pm 2\sqrt{2}{\rm Im}(\beta)
e^{-\Gamma\tau}-2iP_z\,.
\end{eqnarray}

The Wigner function (\ref{Wsideth}) is plotted in
Fig.6 and shows that the cat state (\ref{cat}) 
is very sensitive to the noise induced by the
measurement.

\section{Conclusions}

In conclusion, we have studied a trapped electron 
interacting with a standing radiation field and have
shown that several interesting features can arise
when the dipole approximation is not invoked. 
First, the proposed model provide a method for indirect 
measurement on the cyclotron degree of freedom.
On the other hand, the possibilities to generate 
nonclassical states could be useful to test
the linearity of Quantum Mechanics \cite{wei},
as well as to probe the decoherence of a mesoscopic
system \cite{zur}.
Furthermore, it is worth noting that the entanglement 
induced by the 
radiation field 
could also be used to explore the quantum logic 
possibilities of a trapped electron system.  

Hence, the geonium system in such configuration 
could result alternative and/or complementary 
to other trapped systems. In addition it has the 
advantage of involving a structureless particle, 
while for example an ion in a Paul trap behaves
as a two-level system only ideally.
Moreover, having the electron an antiparticle,
the model could also be  used to perform some 
fundamental tests of simmetry.

Finally, based on theese considerations we conclude that 
it should be an interesting challenge to 
experimentally implement this model.
The realistic values of parameters (see e.g. Ref. \cite{BG}) 
we have used yield that feasible with the actual technology.
The main problem could be represented by the low values of 
$N_{th}$ in Sec. IV, however, to better evidenciate the 
desired effects one 
could adjust the 
experimental set up in order to increase the  
dishomegeneity of the field experienced by the particle
(to this end, we note that traps bigger than the usual 
are available as well \cite{cern}).

\section*{Appendix A}

We consider the position space matrix 
elements 
of the state ${\hat\rho}(t)$, i.e.
\begin{equation}\label{rhoel}
{}_c\langle Q+Y|\,{}_z\langle Z'|{\hat\rho}(t)
|Z''\rangle_z\,|Q-Y\rangle_c\,,
\end{equation}
and we denote them as $\wp(Z',Z'')$ since the evolution 
will take place only in the axial space.
The dependence on the cyclotron  variables 
remains implicit.
Then, with the aid of Eq. (\ref{Psit}) we get
\begin{equation}\label{Pt}
\wp(Z',Z'')=\sum_{m,n}C_mC_n^*\exp\left[-\left({Z'}^2
+{Z''}^2\right)/2\right]
H_m\left(Z'\right)
H_n\left(Z''\right)\,,
\end{equation}
where $H_m$ are the Hermite polynomials, and
\begin{eqnarray}\label{CCst}
C_m&=&\left(\frac{1}{\pi}\right)^{1/4}
\frac{1}{\sqrt{2^mm!}}
e^{-|\beta|^2/2}\frac{\beta^m}{\sqrt{m!}}
\langle Q+Y|\zeta_m \rangle\,,\\
C_n^*&=&\left(\frac{1}{\pi}\right)^{1/4}
\frac{1}{\sqrt{2^nn!}}
e^{-|\beta|^2/2}\frac{(\beta^*)^n}{\sqrt{n!}}
\langle \zeta_n|Q-Y\rangle\,.
\end{eqnarray}
The master equation for the free evolution 
in a thermal bath \cite{gar} has the corresponding 
partial differential equation for the probability $\wp$
\begin{eqnarray}\label{meth1}
\partial_{\tau}\wp(Z',Z'',\tau)&=&\Bigg\{\frac{i}{2}
\left(\frac{\partial^2}{\partial {Z'}^2}
-\frac{\partial^2}{\partial {Z''}^2}\right)
-\frac{\Gamma}{2}(Z'-Z'')
\left(\frac{\partial}{\partial Z'}
-\frac{\partial}{\partial Z''}\right)\nonumber\\
&-&\Gamma N_{th}(Z'-Z'')^2
\Bigg\}\wp(Z',Z'',\tau)\,,
\end{eqnarray}
where we have set $\Gamma=\gamma_z/m_0$. Both, 
the damping constant and the time are scaled by
the axial frequency, i.e. $\Gamma/\omega_z\to\Gamma$ 
and $\tau\omega_z\to\tau$. 

The differential equation (\ref{meth1}) is 
considerably simplified by the change of variables
\begin{eqnarray}
Z'&=&u+v\,,\label{xuz}\\
Z''&=&u-v\,,\label{yuz}
\end{eqnarray}
leading to 
\begin{equation}\label{meth2}
\partial_{\tau}\wp(u,v,\tau)=
\left\{\frac{i}{2}\frac{\partial^2}{\partial u\partial v}
-\Gamma v\frac{\partial}{\partial v}
-4\Gamma N_{th}v^2
\right\}\wp(u,v,\tau)\,.
\end{equation}
By using the Fourier transform
\begin{equation}\label{FTP}
\wp(u,v)=\int dq\,e^{2iqu}{\tilde \wp}(q,v)\,,
\end{equation}
Eq. (\ref{meth2}) becomes
\begin{equation}\label{meth3}
\frac{\partial{\tilde \wp}}{\partial\tau}+
\left(q+\Gamma v\right)
\frac{\partial{\tilde \wp}}{\partial v}
=-4\Gamma N_{th}v^2{\tilde \wp}\,,
\end{equation}
which can be solved by the method of characteristics.
The solution takes the form 
\begin{eqnarray}\label{methsol}
{\tilde \wp}(q,v,\tau)&=&{\tilde \wp}\left(q,
\left[\left(v+\frac{q}{\Gamma}\right)
e^{-\Gamma\tau}
-\frac{q}{\Gamma}\right],0\right)\nonumber\\
&\times&\exp\Bigg\{
-2N_{th}\left(v+\frac{q}{\Gamma}\right)^2
\left(1-e^{-2\Gamma\tau}\right)\nonumber\\
&+&\frac{8N_{th}}{\Gamma}q\left(v+\frac{q}{\Gamma}\right)
\left(1-e^{-\Gamma\tau}\right)\Bigg\}
e^{-4q^2N_{th}\tau/\Gamma}\,.
\end{eqnarray}

In our case, from the Eqs. (\ref{Pt}), (\ref{xuz}), 
(\ref{yuz}), (\ref{FTP}), results
\begin{eqnarray}\label{Ptilini}
{\tilde \wp}(q,v,0)&=&\sqrt{\pi}
\exp\left[-v^2-q^2\right]\nonumber\\
&\times&\Bigg\{\sum_{m<n}C_mC_n^*2^nm!\left(-v
-iq\right)^{n-m}
L^{n-m}_m\left[2\left(v^2
+q^2\right)\right]\nonumber\\
&+&\sum_{m=n}|C_n|^22^nn!
L_m\left[2\left(v^2
+q^2\right)\right]\nonumber\\
&+&\sum_{m>n}C_mC_n^*2^mn!\left(-v
-iq\right)^{m-n}
L^{m-n}_n\left[2\left(v^2
+q^2\right)\right]\Bigg\}\,,
\end{eqnarray}
where $L^m_n$ indicates the associated Laguerre 
polynomials.
Therefore, starting from the above expression, 
the solution (\ref{methsol}) can be easily
constructed.

The Wigner function of the 
cyclotron state after a measurement giving the result
$p_z$ (or equivalently $P_z$),  will be
\begin{equation}\label{Wth1}
W(Q,P,\tau)={\cal N}^{\,2}\,\int dY\,{}_c\langle 
Q+Y|\,{}_z\langle P_z|{\hat\rho}(t+\tau)
|P_z\rangle_z\,|Q-Y\rangle_c \,e^{-2iPY}\,,
\end{equation}
where ${\cal N}$ is the normalization constant needed 
after the projection.
By inserting identities in terms of the set of states 
$\{|u\pm v\rangle_z\}$,
and with the aid of the Fourier transform (\ref{FTP}),
we get 
\begin{equation}\label{Wth2}
W(Q,P,\tau)={\cal N}^{\,2}\,\int dY\,\int dv\,
{\tilde \wp}(0,v,\tau)e^{-2ivP_z-2iYP}\,.
\end{equation}
The dependence on the 
cyclotron variables $Q$ and $Y$ is implicit on
$\wp$.
Hence, by performing the integration one arrives 
at the expression (\ref{Wth3}).

\section*{Appendix B}

If $\tau$ is the duration of the measurement, at the end of the 
measurement we have
\begin{eqnarray}
W(Z,P_z,\tau)&=&\frac{1}{\pi}\int dv\,\langle Z
+v|{\hat\rho}_z(\tau)|Z-v\rangle e^{-2iP_zv}\,,\\
&=&\frac{1}{\pi}\int dv\, \wp(Z,v,\tau) e^{-2iP_zv}\,,\\
&=&\frac{1}{\pi^2}\int dv\,\int dq\, 
{\tilde \wp}(q,v,\tau) e^{-2iP_zv+2iqZ}\label{W11}\,,
\end{eqnarray}
where ${\tilde \wp}(q,v,\tau)$ is the same of 
Eq. (\ref{methsol}), but with the initial condition
determined by Eq.(\ref{cat})
\begin{eqnarray}\label{Ptilini2}
{\tilde \wp}(q,v,0)&=&\frac{1}{2}e^{-|\beta|^2
-q^2-v^2+\beta^2/2+\beta^{*2}/2}\nonumber\\
&\times&\Bigg\{
\exp\left[2{\rm Im}(\beta)^2-2\sqrt{2}i{\rm Im}(\beta)q
-2\sqrt{2}i{\rm Re}(\beta)v\right]
\nonumber\\
&+&\exp\left[2{\rm Im}(\beta)^2+2\sqrt{2}
i{\rm Im}(\beta)q+2\sqrt{2}i{\rm Re}(\beta)v\right]
\nonumber\\  
&-&i\exp\left[-2{\rm Re}(\beta)^2
-2\sqrt{2}{\rm Re}(\beta)q+2\sqrt{2}{\rm Im}(\beta)v\right]
\nonumber\\ 
&+&i\exp\left[-2{\rm Re}(\beta)^2
+2\sqrt{2}{\rm Re}(\beta)q-2\sqrt{2}{\rm
Im}(\beta)v\right]\Bigg\}\,.
\end{eqnarray}
Thus, performing the double integral in Eq.(\ref{W11}) we get 
the expression (\ref{Wsideth}).

\section*{Acknowledgments}
A.M.M. would like to thank the University 
of Camerino for the kind 
hospitality.
S.M. would like to thank the Instituto 
Superior T\'ecnico for the kind  
hospitality and the INFN (Sezione Perugia)
for partial support.

\newpage

FIGURE CAPTIONS

Fig.1 Spectrum of axial momentum for $\Delta=1.5\times10^4$ 
${\rm s}^{-1}$,
$\kappa^2=10^{-6}$, $\gamma_c=1.5$ ${\rm s}^{-1}$,
$\gamma_z/m_0=20$ ${\rm s}^{-1}$,
$|\epsilon|=1.4\times 10^4$, ${\rm s}^{-1}$, 
$\varphi=3\pi/4$,
$f=10^{11}$ ${\rm s}^{-1}$,
and
$N_{th}=10^3$.
The peak on the right represents the resonance in                  
absence of coupling.
The separation between peaks is proportional to 
the cyclotron quadrature 
amplitude.

Fig.2 Variance for the cyclotron 
quadratures
$X_{\varphi=0}$ (dashed line) and $X_{\varphi=\pi/2}$ 
(solid line) as function of the detuning $\Delta$.
The values of other parameters are the same of Fig.1.

Fig.3 The Wigner function of Eq. (\ref{Wig}) is plotted 
for the parameters $\beta=1$, $\epsilon\kappa^2t=-2.4i$,
after an axial momentum 
measurement yielding the most probable value of $p_z$. 

Fig.4 The same of Fig.3 including the effects of finite 
time measurement.
Here $\Gamma\tau=0.1$ and $N_{th}=10$.

Fig.5 The Wigner function of cat state (\ref{cat}) 
is plotted for $\beta=2$.

Fig.6 The same of Fig.5 including the effects of 
finite time measurement.
Here  $\Gamma=6$, $\tau=0.4$, and $N_{th}=10$.

\end{document}